\documentclass[a4paper]{VisionStyle}
\usepackage{epsfig}

\begin{document}
\def\be{\begin{equation}}
\def\ee{\end{equation}}

\title{Unifying MeV-blazars with GeV-blazars}

\author{M.\,B{\l}a{\.z}ejowski\inst{1} \and M.\,Sikora\inst{1,2} \and 
  R.\,Moderski\inst{1} \and G.\,Madejski\inst{2} } 

\institute{
 Nicolaus Copernicus Astronomical Center, Warsaw, POLAND 
\and 
 Stanford Linear Accelerator Center
Menlo Park, CA 94025, USA
 }

\maketitle 

\begin{abstract}
We demonstrate that the spectral differences between Flat Spectrum Radio Quasars 
(FSRQ) with 
steep gamma-ray spectra (MeV-blazars) and  FSRQ with flat gamma-ray spectra
(GeV-blazars)
can be explained by assuming that in the MeV-blazars, the production of 
gamma-rays is dominated by Comptonization of infrared radiation of 
hot dust, whereas in the GeV-blazars --- by Comptonization of broad 
emission lines.  Additional ingredient, required to reach satisfactory
unification, is an assumption that the radiating electrons are accelerated via 
a two step process, in the lower energy range -- following instabilities 
driven by shock-reflected ions, and in the higher energy range --
via resonant scatterings off Alfven waves.
Our model predicts that on average, the MeV-blazars should vary on 
longer time scales than GeV-blazars.

\end{abstract}

\section{Assumptions}
\label{mblazejowski-C2_sec:as}
  
\noindent

\begin{itemize}
\item  Relativistic electrons  in a subparsec jet are produced 
in shocks formed by colliding inhomogeneities. The inhomogeneities 
are assumed to be intrinsically identical and to move down the jet
with bulk Lorentz factors $\Gamma_2 >\Gamma_1 \gg 1$;

\item Time scale of the collision,
as measured in the comoving frame of the shocked plasma, is
\be t_{coll}' \simeq t_{fl} {\cal D}  \, ,\ee
where $t_{fl}$ is the observed time scale of the flare, and
\be {\cal D} \equiv {1 \over \Gamma (1 - \beta \cos{\theta_{obs}})} \ee
is the Doppler factor of the shocked plasma, and 
\be \Gamma \simeq \sqrt {\Gamma_1 \Gamma_2} \, ; \ee 

\item Inertia of inhomogeneities is dominated  by protons
(i.e. $n_e/n_p \ll m_p/m_e$);

\item Efficiency of energy dissipation is defined as 
\be \eta_{diss} \simeq 
{((\Gamma_2/\Gamma_1)^{1/2} - 1)^2 \over (\Gamma_2/\Gamma_1)+1} \, ,\ee

\item 
Injection of relativistic electrons is approximated by a two-power-law 
function, with the break at $\gamma_b$, at which magnetic rigidity
of electrons is equal to rigidity of thermal protons, i.e., when their 
momenta are equal 
\be m_e \sqrt{\gamma_b^2 -1} \simeq  m_p \sqrt {\gamma_{p,th}^2-1} \, ,\ee
where 
\be \gamma_{p,th} -1 = \eta_{p,th} \kappa  \ee
is the average thermal  proton energy in the shocked
plasma, $\eta_{p,th}$  is the fraction of the dissipated  energy 
tapped to heat the protons and \be \kappa \simeq
{ ((\Gamma_2/\Gamma_1)^{1/2} - 1)^2 \over 2(\Gamma_2/\Gamma_1)^{1/2}} \, .  \ee
 is the amount of energy dissipated per proton in units of $m_pc^2$. 
\noindent
\item  Since the time scales of the flares in FSRQ are rarely shorter than 1 day, 
the distances of their production, 
\be r_{fl} \sim (r_{fl} /\Delta r_{coll}) c t_{fl} {\cal D} \Gamma \ee
are expected to be larger than 0.1 parsec (where $\Delta r_{coll}$ 
is a distance range over which the flare is produced). At such distances,
the largest contribution to the energy density of an external radiation 
field $u_{ext}'$ is provided by the diffuse component of the 
broad emission lines and infrared radiation of hot dust.
\end{itemize}

\section{Spectra}
\label{mblazejowski-C2_sec:sp}
The basic feature of the high energy spectra in FSRQ --- 
a break between the  $\gamma$-ray and  the X-ray spectral portions ---
has a natural explanation in terms of the External Radiation Comptonization 
(ERC) model (\cite{mblazejowski-C2:sik94}). In this model,
X-ray spectra are produced by electrons with
radiative cooling time scale, $t_{cool}'$, longer than the 
collision time scale, $t_{coll}'$
({\it slow cooling regime}), whereas $\gamma$-rays are produced by 
electrons with $t_{cool}' < t_{coll}'$ ({\it fast cooling regime}).  
Noting that cooling rate  of electrons, dominated by Comptonization
of external radiation, is
\be \dot \gamma \simeq { c \sigma_T  \over m_e c^2} u_{ext}' \gamma^2  \ee
we obtain that cooling time scale is
\be t_{cool}' \simeq {\gamma \over \dot \gamma} \simeq
{m_e c^2 \over  c \sigma_T}{1 \over  \gamma u_{ext}'} \ee
where $u_{ext}'$ is the energy density of an external radiation field.
Then, from $t_{cool}' = t_{coll}'$, where $t_{coll}'$ is given by Eq. (1),
the break  in the electron distribution is
\be \gamma_c \simeq {m_e c^2 \over c \sigma_T} 
{1 \over u_{ext}' t_{fl} \Gamma^2} \, \ee
For $\gamma < \gamma_c$, the slope of the electron distribution is
the same as the slope of the injection function, while for $\gamma > \gamma_c$,
the slope of the electron energy distribution is steeper by $\Delta s=1$
($s: N_{\gamma} \propto \gamma^{-s}$). 

For  $u_{ext}'  \simeq u_{diff} \Gamma^2$  ,  
we predict  that  the break in an electron energy distribution at $\gamma_c$ 
should be imprinted in the electromagnetic spectrum at frequency
\be \nu_c \sim \Gamma^2 \gamma_c^2 \nu_{diff}
\simeq \big({m_e c^2 \over \sigma_T}\big)^2 
{\nu_{diff} \over u_{diff}^2 \Delta r_{coll}^2 }, \ee
where $\nu_{diff}$ is the characteristic frequency of the diffuse 
external radiation field.

\begin{figure}[ht]
  \begin{center}
    \epsfig{file=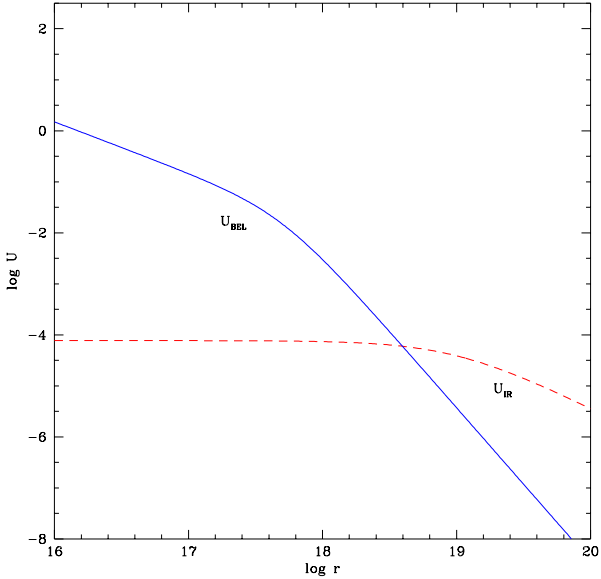, width=8.5cm}
  \end{center}
\caption{ The dependence of energy density of $u_{BEL}$ (solid line)
 and $u_{IR}$ (dashed line) on distance from the central source is 
shown. }  
\label{mblazejowski-C2_fig:fig1}
\end{figure}

The predicted location of the break
does not depend on $\Gamma$, and, 
for typical energy densities of BELR and
of the hot dust  radiation, is comparable to the 
range $10^{20} - 10^{22}$Hz determined from observations (see Fig. 2).
The presented results are obtained using the following 
approximations for energy densities of the diffused external radiation components:
\be u_{BEL,diff}(r) \simeq 
{\xi_{BEL}(r_{BEL}) L_{UV} \over 4 \pi r^2  
((r/r_{BEL}) + (r_{BEL}/r))c}  \ee
where
\be r_{BEL} \sim 3 \times 10^{17} \sqrt{L_{UV,46}} {\rm cm}  \ee
and 
\be u_{IR,diff}(r) \sim {\xi_{IR}(r_{d,min}) L_{UV}  
\over 4 \pi (r_{d,min}^2  + r^2 (r_{d,min}/r)^{-0.68})c}
\ee
where 
\be r_{d,min} \simeq 
\big ({ L_{UV} \over 4 \pi \sigma_{SB} T_{d,max}^4} \big)^{1/2}  \ee
is the minimum distance of dust to the central source and 
$T_{d,max}$ is the maximum temperature of dust.

\begin{figure}[ht]
  \begin{center}
    \epsfig{file=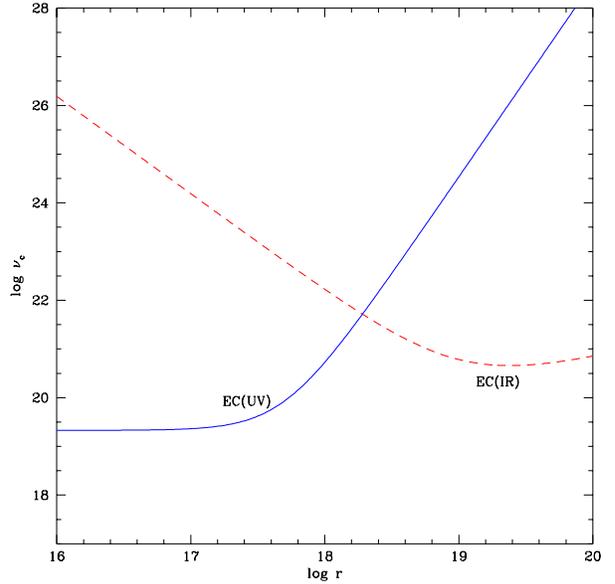, width=8.5cm}
  \end{center}
\caption{ The dependence of $\nu_c$ on the distance.  
Solid line- $\nu_{EC(UV),c}$, dashed line- $\nu_{EC(IR),c}$. }  
\label{mblazejowski-C2_fig:fig2}
\end{figure}

Due to the cooling effect the spectra should have a break by $\Delta \alpha = 0.5$, 
and this is consistent with the spectral breaks observed in $\gamma$-ray
quasars during flares.
In these quasars, the spectral index in the EGRET range is 
$\alpha_{\gamma} \le 1$ (\cite{mblazejowski-C2:pohl97}). 
There are, however,  several quasars, called MeV-blazars, which  have  
the spectral break $\Delta \alpha$ much larger than $0.5$.
Their $\gamma$-ray spectra are very soft, with $\alpha_{\gamma} \ge 1.5$, 
and the X-ray spectra are very hard, with $\alpha_X \le 0.5$ 
(\cite{mblazejowski-C2:tav00}). 
In order to explain such spectra, it is necessary to postulate 
the break in the electron injection function. Origin of the break can be 
related to the two-step acceleration process of electrons and its location
is likely to be at $\gamma_b$ (see Eq.(5)).
For $2.5<\Gamma_2 / \Gamma_1 <10$ and $\eta_{p,th}=0.5$, the break in 
the electron injection function is $600<\gamma_b<1700$. 

\begin{figure}[ht]
  \begin{center}
    \epsfig{file=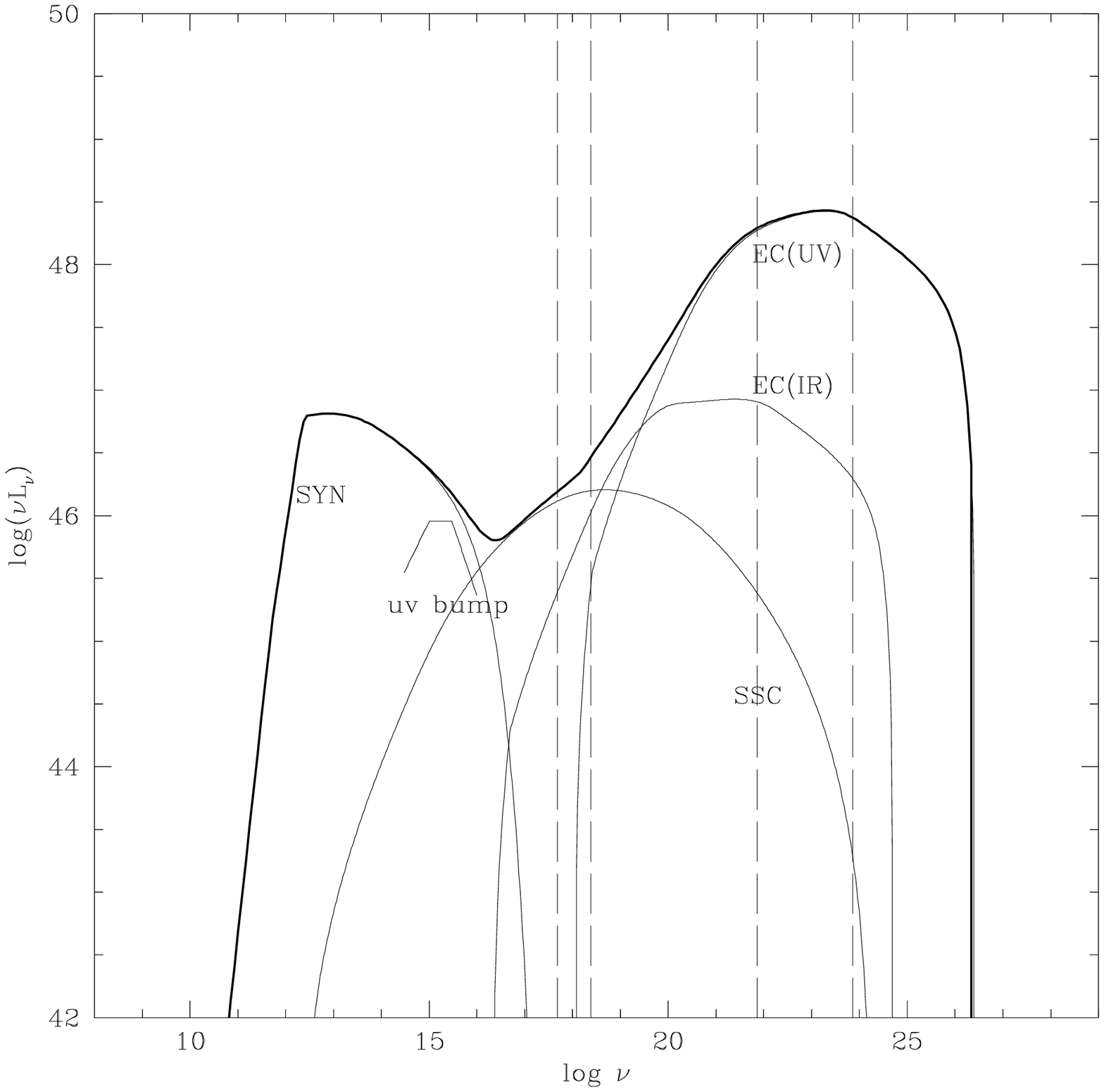, width=8.5cm}
  \end{center}
\caption{ The broadband  spectrum of a ``typical'' $GeV$-blazar (for
the justification of the parameters used in the model, 
see Sikora et al. 2002 in preparation).  The two areas confined 
by vertical dashed lines represent $2-10$ keV and 30 MeV - 3 GeV bands respectively. }  
\label{mblazejowski-C2_fig:fig3}
\end{figure}

External photons scattered by electrons with Lorentz factors $\gamma_b$ 
are boosted up to energies
\be \nu_b \sim \gamma_b^2 \Gamma^2 \nu_{ext} \ee
Hence, the break in the electron injection function is imprinted 
in the electromagnetic spectrum at
around $1$ GeV if $\nu_{ext} =\nu_{BEL}$, and at ten times lower
energies, if $\nu_{ext} =\nu_{IR}$.
In the former case $\nu_b \gg \nu_c$ and the $\gamma$-ray spectra
in the EGRET range have slopes $\alpha_{\gamma} \le 1$ (see Fig.(3)), 
while in the latter case $\nu_b \sim \nu_c \le 30$ MeV, and the 
$\gamma$-ray spectra in the EGRET range are very soft (see Fig.(4)). 

The above scheme can also explain two other differences between 
MeV-blazars and GeV-blazars. One is that in MeV-blazars, the contribution
to the spectra from  thermal UV-bump is often more apparent than in 
GeV-blazars (see, e.g., \cite{mblazejowski-C2:tav00}), and the other is that 
the X-ray spectra are harder in MeV-blazars than in GeV blazars.
The former results in our model from the fact that due to the decrease 
of magnetic field with distance, the break in the electron
synchrotron spectrum at $\nu_{syn,b}$ is shifted to lower frequencies
at larger distances, and the dilution of thermal UV radiation by synchrotron
component is reduced.  The latter is a result of the fact that the 
SSC spectrum is produced at larger distances and thus it is shifted to lower 
frequencies and has lower intensity;  furthermore, at those distances, 
the dust covering factor is larger than the BELR covering factor.  

Additional prediction of our scheme is that time scales of flares produced 
in MeV blazars should on average be longer than the time scales of flares 
produced in GeV blazars. With the planned sensitive gamma-ray observatories 
such as GLAST, this prediction can be verified in future 
statistically, but also via studies of individual objects such as 
PKS 0208-512, which sometimes show very hard $\gamma$-ray spectra and 
on other occasions --- very soft $\gamma$-ray spectra 
(\cite{mblazejowski-C2:blom96}, \cite{mblazejowski-C2:stac96}).

\begin{figure}[ht]
  \begin{center}
    \epsfig{file=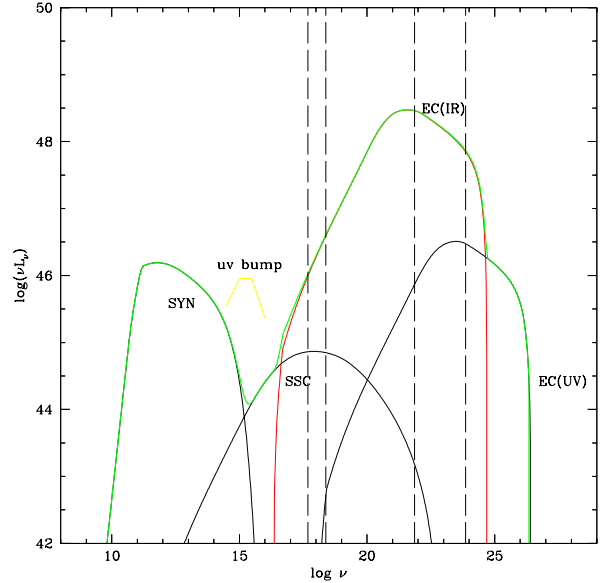, width=8.5cm}
  \end{center}
\caption{  The  broadband spectrum of a ``typical'' $MeV$-blazar. }  
\label{mblazejowski-C2_fig:fig4}
\end{figure}

\begin{acknowledgements}
This work was partially supported by the Polish Committee of Scientific
Research grant no. 5 P03D 002 21, and by NASA Chandra grant GO0 - 1038A
to Stanford University.  
\end{acknowledgements}

\end{document}